\documentclass[conference]{IEEEtran}

\IEEEoverridecommandlockouts

\usepackage{graphicx}
\usepackage{amsmath}
\usepackage{amssymb}
\usepackage{color}

\usepackage{algorithm}
\usepackage{algpseudocode}
\algnewcommand\algorithmicinput{\textbf{Input:}}
\algnewcommand\INPUT{\item[\algorithmicinput]}
\algnewcommand\algorithmicoutput{\textbf{Output:}}
\algnewcommand\OUTPUT{\item[\algorithmicoutput]}

\usepackage{mathtools}
\DeclarePairedDelimiter{\ceil}{\lceil}{\rceil}
\DeclarePairedDelimiter{\floor}{\lfloor}{\rfloor}

\graphicspath{{figures/}}

\newtheorem{definition}{Definition}
\newtheorem{lemma}{Lemma}
\newtheorem{theorem}{Theorem}
\newtheorem{corollary}{Corollary}

\newtheorem{example}{Example}

\newcommand{\F}{\mathbb{F}}

\newcommand{\remove}[1]{}

\newcommand\nc\newcommand
\nc\bfa{{\boldsymbol a}}\nc\bfA{{\bf A}}\nc\cA{{\mathcal A}}
\nc\bfb{{\boldsymbol b}}\nc\bfB{{\bf B}}\nc\cB{{\mathcal B}}
\nc\bfc{{\boldsymbol c}}\nc\bfC{{\bf C}}\nc\cC{{\mathcal C}}
\nc\bfd{{\boldsymbol d}}\nc\bfD{{\bf D}}\nc\cD{{\mathcal D}}
\nc\bfe{{\boldsymbol e}}\nc\bfE{{\bf E}}\nc\cE{{\mathcal E}}
\nc\bff{{\boldsymbol f}}\nc\bfF{{\bf F}}\nc\cF{{\mathcal F}}
\nc\bfg{{\boldsymbol g}}\nc\bfG{{\bf G}}\nc\cG{{\mathcal G}}
\nc\bfh{{\boldsymbol h}}\nc\bfH{{\bf H}}\nc\cH{{\mathcal H}}
\nc\bfi{{\boldsymbol i}}\nc\bfI{{\bf I}}\nc\cI{{\mathcal I}}
\nc\bfj{{\boldsymbol j}}\nc\bfJ{{\bf J}}\nc\cJ{{\mathcal J}}
\nc\bfk{{\boldsymbol k}}\nc\bfK{{\bf K}}\nc\cK{{\mathcal K}}
\nc\bfl{{\boldsymbol l}}\nc\bfL{{\bf L}}\nc\cL{{\mathcal L}}
\nc\bfm{{\boldsymbol m}}\nc\bfM{{\bf M}}\nc\cM{{\mathcal M}}
\nc\bfn{{\boldsymbol n}}\nc\bfN{{\bf N}}\nc\cN{{\mathcal N}}
\nc\bfo{{\boldsymbol o}}\nc\bfO{{\bf O}}\nc\cO{{\mathcal O}}
\nc\bfp{{\boldsymbol p}}\nc\bfP{{\bf P}}\nc\cP{{\mathcal P}}
\nc\bfq{{\boldsymbol q}}\nc\bfQ{{\bf Q}}\nc\cQ{{\mathcal Q}}
\nc\bfr{{\boldsymbol r}}\nc\bfR{{\bf R}}\nc\cR{{\mathcal R}}
\nc\bfs{{\boldsymbol s}}\nc\bfS{{\bf S}}\nc\cS{{\mathcal S}}
\nc\bft{{\boldsymbol t}}\nc\bfT{{\bf T}}\nc\cT{{\mathcal T}}
\nc\bfu{{\boldsymbol u}}\nc\bfU{{\bf U}}\nc\cU{{\mathcal U}}
\nc\bfv{{\boldsymbol v}}\nc\bfV{{\bf V}}\nc\cV{{\mathcal V}}
\nc\bfw{{\boldsymbol w}}\nc\bfW{{\bf W}}\nc\cW{{\mathcal W}}
\nc\bfx{{\boldsymbol x}}\nc\bfX{{\bf X}}\nc\cX{{\mathcal X}}
\nc\bfy{{\boldsymbol y}}\nc\bfY{{\bf Y}}\nc\cY{{\mathcal Y}}
\nc\bfz{{\boldsymbol z}}\nc\bfZ{{\bf Z}}\nc\cZ{{\mathcal Z}}
\nc\od{{\bar d}}\nc\ow{{\bar w}}\nc\odelta{{\bar\delta}}
\nc\ox{{\bar x}}\nc\oy{{\bar y}}\nc\ou{{\bar u}}
\nc\oh{{\bar h}}
\DeclareMathOperator{\E}{\text{\sf E}}

\begin{document}
\title{On One Generalization of LRC Codes with Availability}
\date{}

\author{
  \IEEEauthorblockN{Stanislav Kruglik\IEEEauthorrefmark{1}\IEEEauthorrefmark{2}\IEEEauthorrefmark{3}, Marina Dudina\IEEEauthorrefmark{1}, Valeriya Potapova\IEEEauthorrefmark{1}\IEEEauthorrefmark{2} and Alexey Frolov\IEEEauthorrefmark{1}\IEEEauthorrefmark{2}}
	
 \IEEEauthorblockA{\small \IEEEauthorrefmark{1} Skolkovo Institute of Science and Technology\\
    Moscow, Russia
    }
 \IEEEauthorblockA{\small \IEEEauthorrefmark{2} Institute for Information Transmission Problems\\
    Russian Academy of Sciences\\Moscow, Russia
    }
 \IEEEauthorblockA{\small \IEEEauthorrefmark{3} Moscow Institute of Physics and Technology \\
    Moscow, Russia
    }

  {\small stanislav.kruglik@skolkovotech.ru, marina.dudina@skolkovotech.ru,  valeriya.potapova@skolkovotech.ru, al.frolov@skoltech.ru}
      
}

\maketitle

\begin{abstract}
We investigate one possible generalization of locally recoverable codes (LRC) with all-symbol locality and availability when recovering sets can intersect in a small number of coordinates. This feature allows us to increase the achievable code rate and still meet load balancing requirements. In this paper we derive an upper bound for the rate of such codes and give explicit constructions of codes with such a property. These constructions utilize LRC codes developed by Wang et al.
\end{abstract}

\section{Introduction}

A locally recoverable code (LRC) is a code over finite alphabet such that each symbol is a function of small number of other symbols that form a recovering set \cite{Gopp11,Gopp14,Papailiopoulos,Rawat,Yekhanin}. These codes are important due to their applications in distributed and cloud storage systems. LRC codes are well-investigated in the literature. The bounds on the rate and minimum code distance are given in \cite{Gopp11, Papailiopoulos} for the case of large alphabet size. The alphabet-dependent shortening bound (see \cite{Litsyn} for the method explanation) is proposed in \cite{Mazumdar}. Optimal code constructions are given in \cite{Silberstein} based on rank-metric codes (for large alphabet size, which is an exponential function of the code length) and in \cite{TamoBarg} based on Reed-Solomon codes (for small alphabet, which is a linear function of the code length).       

The natural generalization of an LRC code is an LRC code with availability (or multiple disjoint recovering sets).  Availability allows us to handle multiple simultaneous requests to erased symbol in parallel. This property is very important for hot data that is simultaneously requested by a large number of users. The case of LRC codes with availability is much less investigated. Bounds on parameters of such codes and constructions are given in \cite{Rawat, TamoBargFrolov, Parakash, Yaakobi}. Most of the papers focused on information-symbol locality and availability.  In what follows we are interested in all-symbol locality and availability that is preferable in applications as it permits a uniform approach system design. 



The property of availability decreases maximum achievable code rate \cite{TamoBargFrolov}. In this paper we propose a new generalization of LRC codes with availability. Namely, we assume that recovering sets can intersect in a small number of coordinates. This feature allows us to increase the achievable code rate and still meet load balancing requirements.

Our contribution is as follows. We investigate one possible generalization of locally recoverable codes (LRC) with all-symbol locality and availability when recovering sets can intersect in a small number of coordinates. We derive an upper bound for the rate of such codes and give explicit constructions of codes with such a property. These constructions utilize LRC codes developed in \cite{WangZhang}.

\section{Preliminaries}

\subsection{LRC codes}
Let us denote by $\F_q$ a field with $q$ elements. Let $[n] = \{1, 2, \ldots, n\}$. The code $\cC \subset \F_q^n$ has locality $r$ if every symbol of the codeword $c\in \cC$ can be recovered from a subset of $r$ other symbols of $c$  \cite{Gopp11}. In other words, this means that, given $c\in \cC, i\in [n],$ there exists a subset of coordinates 
${\cR}_i\subset [n]\backslash i, |{\cR}_i|\le r$ such that the restriction of $\cC$ to the coordinates in ${\cR}_i$ enables one to find the value of $c_i.$ The subset ${\cR}_i$ is called a {\em recovering set} for the symbol $c_i$.

\subsection{LRC codes with availability}

Generalizing this concept, assume that every symbol of the code $\cC$ can be recovered from $t$ disjoint subsets of symbols of size $r$. More formally, denote by $\cC_I$ the restriction of the code $\cC$ to a subset of coordinates $I\subset [n]$.
Given $a\in \F_q$ define the set of codewords
   $
   \cC(i,a)=\{c\in \cC: c_i=a\},\; i\in[n].
   $
   
\begin{definition}
A code $\cC$  is said to have $t$ disjoint recovering sets if for every $i \in [n]$ there are $t$ pairwise disjoint subsets ${\cR}_{i}^1,\dots,{\cR}_{i}^t\subset [n]\backslash i$ such that for all $j=1,\dots,t$ and every pair of symbols $a,a'\in \F_q, a\ne a'$
  \begin{equation*}\
  \cC(i,a)_{{\cR}_{i}^j}\cap \cC(i,a')_{{\cR}_{i}^j}=\emptyset.
  \end{equation*}
  \label{def:tlrc}
\end{definition}

In what follows we refer these codes as $(r,t)$-LRC codes. We briefly list the existing results below. The first bound for $(r,t)$-LRC codes was given in \cite{Wang2, Rawat2}
 \[
 d \leq n-k+2-\ceil*{\frac{t(k-1)+1}{t(r-1)+1}}.
 \]
  
An improvement of this bound was obtained in \cite{TamoBargFrolov}
\begin{equation*}\label{Barg}
  d \leq n - \sum_{i=0}^t \floor*{\frac{k-1}{r^i}}.
\end{equation*}
 
 An alphabet-dependent bound was proposed in \cite{Yaakobi} and has form 
  \begin{equation*}\label{Yaakobi}
  d \leq \min\limits_{\begin{subarray}{c}
 1\leq x \leq \ceil*{\frac{k-1}{(r-1)t+1}};{}{} 1 \leq y_j \leq t;{}{} j \in [x] \\ A<k;{}{}x,y_j\in Z^{+}
 \end{subarray}} d_{l-opt}^{q}[n-B,k-A] , 
  \end{equation*}
 where $A=\sum_{j=1}^x (r-1)y_j+x$, $B=\sum_{j=1}^x ry_j+x$ and $d_{l-opt}^{q}$ denote the
largest possible minimum distance of a code over $\F_q$.

The bound on the rate of $(r,t)$-LRC codes was given in \cite{TamoBargFrolov}
\begin{equation}\label{eq::rate}
\frac{k}{n} \leq R^*(r,t) = \prod_{i=1}^{t} \frac{1}{1+\frac{1}{ir}}. 
\end{equation}

This bound was improved in \cite{Parakash} for $t=2$.
 
In \cite{WangZhang} a recursive construction of binary $(r,t)$-LRC codes was proposed. The parameters of these codes are as follows: $n=\binom{r+t}{t}$, $R=\frac{r}{r+t}$ and $d = t+1$. We refer these codes as WZL codes. WZL code is defined by its' parity-check matrix. Let $m=r+t$. Let us define matrix $\mathbf{H}(m,t)$ as follows. Each row of $\mathbf{H}(m,t)$ is associated with $(t-1)$-subset of $[m]$ sorted in lexicographical order, each column -- with $t$-subset of $[m]$ also sorted in lexicographical order. In this case the element $(i,j)$ of $\mathbf{H}(m,t)$ is equal to $1$ if $E_i \subseteq F_j$, where $E_i$ is $(t-1)$-subset of $[m]$ associated with $i$-th row and $F_j$ is $t$-subset of $[m]$ associated with $j$-th column. It must be mentioned that $\mathbf{H}(m,t)$ has $\binom{m}{t-1}$ rows and $\binom{m}{t}$ columns and has the following structure:   

$$\mathbf{H}(m,t) = \begin{pmatrix}
\mathbf{H}(m-1,t-1) & 0 \\
I_{\binom{m-1}{t-1}} & \mathbf{H}(m-1,t)
\end{pmatrix},$$ 
where $\mathbf{H}(m,m)=\mathbf{H}(m,1)^{T}=(1,...,1)^{T}$ and $\dim(\mathbf{H}(m,1))=\dim(\mathbf{H}(m,t))=m$.

\subsection{LRC codes with availability and intersection of recovering sets}

Let us give to recovering sets an ability to intersect in at most $x$ positions and define this code as $(r,t,x)$-LRC. More formally, we can say

\begin{definition}
A code $\cC$ is said to be $(r,t,x)$-LRC if for every $i \in [n]$ there are $t$ subsets $R_i^1,...,R_i^t \subset [n] \setminus i$ such, that the following relations follow
\begin{enumerate}
\item  for every pair $l,l' \in [t]$, $l \ne l'$
\[
| R_i^l \cap R_i^{l'} | \leq x;
\]
\item  for all $j=1,\dots,t$ and every pair of symbols $a,a'\in \F_q, a\ne a'$
  \[
  \cC(i,a)_{{\cR}_{i}^j}\cap \cC(i,a')_{{\cR}_{i}^j}=\emptyset.
  \]
\end{enumerate}
 \label{def:xtlrc}
\end{definition}

In what follows we investigate the parameters of such codes.




\section{An upper bound on the rate of $(r,t,x)$-LRC codes} 

\subsection{The recovery graph}
Based on the original idea from \cite{TamoBargFrolov} we represent locally recoverable codes with locality $r$ and availability $t$ as a graph $G$ in the following way. In accordance to the Definition~\ref{def:xtlrc} a coordinate $i$ has $t$ recovering sets $\cR_i^1,...\cR_i^t$,  each of size $r$, where $\cR_i^j\subset [n]\backslash i$. Define a directed graph $G$ as follows. The set of vertices $V=[n]$ corresponds to the set of $n$ coordinates of the LRC code. The ordered pair of vertices $(i,j)$ forms a directed edge $i\to j$ if $j\in {\cR}_i^l$ for some $l\in[t]$. We color the edges of the graph with $t$ distinct colors in order to differentiate between the recovering sets of each coordinate. Note, that as the recovering sets can intersect, then some edges may have several colors. We call $G$ the {\em recovery graph} of the code $\cC.$ 

In what follows we need the following lemma
\begin{lemma}
Let $j \in [t]$ and $s = \min\{ j, \floor{r/x}+1 \}$, then
\[
\frac{(2r-(s-1)x)}{2} s = \underline{N}(r, j, x) \leq \left| \bigcup\limits_{l=1}^{j} \cR_i^l \right| \leq \overline{N}(r, j, x) = jr.
\]
\end{lemma}

\begin{IEEEproof}
The upper bound is trivial and correspond to the case, when recovering sets do not intersect. To prove the lower bound assume, that any two recovering sets intersect in exactly $x$ positions, we have
\begin{eqnarray*}
\left| \bigcup\limits_{l=1}^{t} \cR_i^l \right| &\geq& r + (r-x) + (r-2x) + \ldots + (r-(s-1)x)\\
 &=& \frac{(2r-(s-1)x)}{2} s.
\end{eqnarray*}
\end{IEEEproof}

\begin{corollary}
The out-degree of each vertex $i\in V=V(G)$ is upper bounded with $tr$ and lower bounded with $\underline{N}(r, t, x)$.
\end{corollary}

\subsection{Upper bound on the rate}

The proof is very similar to the proof from \cite{TamoBargFrolov}. For the simplicity of the reader we present the proof here in all the details. Let us introduce the following function
\begin{eqnarray*}
f(r, t, x) &=& \sum_{j=1, j=1 \mod 2}^t \binom{t}{j} \frac{1}{\overline{N}(r, j, x)+1}\\
&-& \sum_{j=1, j=0 \mod 2}^t \binom{t}{j} \frac{1}{\underline{N}(r, j, x)+1}\\
\end{eqnarray*}
The following lemma will be used in the proof.

\begin{lemma} \label{lemma:rfrf} 
There exists a subset of vertices $U\subseteq V$ of size at least 
  \begin{equation*}
  |U|\geq n f(r, t, x),
  \end{equation*}
such that for any $U'\subseteq U$, the induced subgraph $G_{U'}$ on the vertices $U'$ 
has at least one vertex $v\in U'$ such that its set of outgoing edges $\{(v,j),j\in U')\}$ is missing at least one color. 
\end{lemma}

\begin{IEEEproof}
For a given permutation $\tau$ of 
the set of vertices $V=[n]$, we define the coloring of some of the vertices as follows:  
The color $j\in [t]$ is assigned to the vertex $v$ if 
   \begin{equation}
     \tau(v)>\tau(m) \quad\text{for all }m\in {\cR}_v^j.
   \label{eq:dfdf}
   \end{equation}
If this condition is satisfied for several recovering sets ${\cR}_v^j$, the vertex $v$ is assigned any of the colors $j$ corresponding to these sets. 
Finally, if this condition is not satisfied at all, then the vertex $v$ is not colored.

Let $U$ be the set of colored vertices, and consider one of its subsets $U'\subseteq U$.

Let $G_{U'}$ be the induced subgraph on $U'$. We claim that there exists $v\in U'$ such that its set of outgoing edges is missing at least one color in $G_{U'}$. Assume toward a contradiction that every vertex 
of $G_{U'}$ has outgoing edges of all $t$ colors. 
Choose a vertex $v\in U'$ and construct a walk through 
the vertices of $G_{U'}$ according to the following rule. 
If the path constructed so far ends at some vertex with color $j,$ choose one of its outgoing edges also
 colored in $j$ and leave the vertex moving along this edge. 
 By assumption, every vertex has outgoing edges of all $t$ colors, so this process, and hence this path can be extended indefinitely. 
Since the graph $G_{U'}$ is finite, there will be a vertex, call it $v_1,$ that is encountered twice. 
The segment of the path that begins at $v_1$ and returns to it has the form
   $$
   v_1\rightarrow v_2\rightarrow... \rightarrow v_l,
   $$
where $v_1=v_l$. For any $i=1,...,l-1$ the vertex $v_i$ and the edge $(v_i,v_{i+1})$ are colored with 
the same color. Hence by the definition of the set $U$ we conclude that 
$\tau(v_i)>\tau(v_{i+1})$ for all $i=1,\dots, l-1,$ a contradiction. 

In order to show that there exists such a set $U$ of large cardinality, we choose the permutation
$\tau$ randomly and uniformly among all the $n!$ possibilities and compute the expected cardinality 
of the set $U.$

Let $A_{v,j}$ be the event that \eqref{eq:dfdf} holds for the vertex $v$ and the color $j.$ Since $\Pr(A_{v,j})$ does not depend
on $v$, we suppress the subscript $v$, and write
  $$
  \Pr(v\in U)=\Pr( \cup_{j=1}^t A_{j}).
  $$
Let us compute the probability of the event $\cup_{j=1}^tA_j.$
Note that for any set $S\subseteq [t]$ the probability of the event that all the $A_j,j\in S$ occur simultaneously can be estimated as follows
     $$
     \frac{1}{\overline{N}(r, |S|, x)+1} \leq P(\cap_{j\in S}A_j) \leq \frac{1}{\underline{N}(r, |S|, x)+1}.
     $$
Hence by the inclusion exclusion formula we get 
\begin{eqnarray*}
\Pr(\cup_{j=1}^t A_j)&=&\sum_{j=1}^t(-1)^{j-1}\binom{t}{j}P(A_1\cap...\cap A_j)\\
&\geq& \sum_{j=1, j=1 \mod 2}^t \binom{t}{j} \frac{1}{\overline{N}(r, j, x)+1}\\
&-& \sum_{j=1, j=0 \mod 2}^t \binom{t}{j} \frac{1}{\underline{N}(r, j, x)+1}\\
&=& f(r, t, x).
\end{eqnarray*}

Now let $X_v$ be the indicator random variable for the event that $v\in U$, then
\begin{align*}
\E(|U|)&=\sum_{v\in V}\E(X_v)\\
&=\sum_{v\in V}\Pr(v\in U)\\
&=n\Pr( \cup_{j=1}^t A_{j})\\
&\geq n f(r,t,x).
\end{align*}
The proof is completed by observing that there exists at least one choice of $\tau$ for which $|U|\geq \E(|U|).$
\end{IEEEproof}

\begin{theorem}
The rate of an $(r, t, x)$-LRC code $\cC$ satisfies
\begin{equation*}
R(\cC) \leq R^*(r,t,x) = 1-f(r, t, x).
\end{equation*}
\end{theorem}
\begin{IEEEproof}
The colored vertices can be viewed as check symbols as they can be recovered from the rest symbols. Thus, the number of information symbols can be estimated as follows
\[
k \leq n(1-f(r, t, x)).
\]  
\end{IEEEproof}

\section{Lower bounds on the rate of $(r,t,x)$-LRC codes}

In this section we derive a lower bound on the rate of codes with all symbol locality and availability in which recovering sets can intersect. To find a lower bound we propose the following rather simple code construction. In what follows we explain how to construct a parity-check matrix of a linear $(r,t,x)$-LRC code. We start with a parity-check of $(\tilde{r}, \tilde{x})$-WZL code. Let is denote the matrix by $\mathbf{H}_\text{WZL}$. The matrix $\mathbf{H}$ of $(r,t,x)$-LRC code is constructed as follows
\[
\mathbf{H} = \mathbf{H}_\text{WZL} \otimes \underbrace {[11 \ldots 1]}_{x+1},
\]
where $\otimes$ denotes a Kronecker product of matrices.

As a result, we have a matrix of length $n = (x+1) \binom{\tilde r+ \tilde t}{\tilde t}$. It is obvious, that each row of the new matrix will have $(x+1)(\tilde r+1)$ ones and the number of positions, in which two recovering sets intersects is equal to $x$. This construction will have the same availability $t$ as it was for the standard WZL code. Thus, the parameters of the resulting code are as follows
\begin{eqnarray*}
r &=& (\tilde{r} +1)(x+1)-1 \\
t &=& \tilde{t}
\end{eqnarray*}

The matrix $\mathbf{H}$ has exactly the same rank as the matrix $\mathbf{H}_\text{WZL}$, the rank is equal to $\binom{\tilde r+ \tilde t-1}{\tilde t-1}$. Thus, the rate of the resulting code can be calculated as follows
\begin{eqnarray*}
R &=& 1 - \frac{\tilde t}{(\tilde r + \tilde t)(x+1) }\\
&=& 1 - \frac{t}{\left(\frac{r+1}{x+1} -1 + t\right) (x+1) )}\\
&=& 1 - \frac{t}{r +t + (t-1)x} \\
&=& \frac{r + (t-1)x}{r +t + (t-1)x}.
\end{eqnarray*}

\begin{example}
Let us start from an $(2, 2)$-WZL code 
\[
\mathbf{H}_\text{WZL} = \left[ 
\begin{array}{cccccccccccc}
     0  &   0  &   0    & 1    & 1  &   1\\
     0  &   1   &  1   &  0  &   0    & 1\\
     1  &   0   &  1  &   0 &    1&     0\\
     1  &   1 &    0   &  1   &  0   &  0
 \end{array}
 \right]
\]
and construct a parity-check matrix of an $(5, 2, 1)$-LRC code. The matrix has a rate $R = 0.75$ and shown below
\[
\mathbf{H} = \left[ 
\begin{array}{cccccccccccc}
   0    &  0 &     0  &    0  &    0  &    0  &    1  &    1  &    1   &   1  &    1  &    1 \\
   0    &  0 &     1  &   1  &    1  &    1   &   0  &    0   &   0   &   0  &    1  &    1 \\
   1    &  1 &     0  &   0  &    1   &   1   &   0  &    0   &   1   &   1 &     0  &    0 \\
    1   &   1 &     1  &    1 &     0  &    0 &     1  &    1  &    0   &   0 &     0 &     0 
 \end{array}
 \right]
 \]
 \end{example}

\section{Numerical results}

In Table~\ref{t_1} we present the comparison of upper bounds on the rate of $(r,t,x)$-LRC codes for different values of parameter $x$. We see that the value of the upper bound increases with the parameter $x$. In the Table~\ref{t_2} we present the comparison of the code rate obtained by proposed code construction ($x=1$) and the code rate of WZL codes with the same locality and availability. In addition, we include the values of the upper bounds for the code rate from \cite{TamoBargFrolov} and the upper bounds for the code rate proposed in this paper. We see, that e.g. for $r=3$, $t=2$ and $x=1$ the lower bound is tight and it is better, then the upper bound for the case of $r=3$, $t=2$ and $x=0$.

\begin{table}
\caption{Upper bounds on the rate of $(r,t,x)$-LRC codes}
\label{t_1}
\centering
\begin{tabular}{| l || c | c | c | c | c |}
  \hline                       
  $(r, t)$ &  $x=0$ & $x=1$ & $x=2$ & $x=3$  \\
  \hline
  (4, 2)   & 0.7111 & 0.7250 & 0.7429 & 0.7667 \\
  (5, 2)   & 0.7576 & 0.7667 & 0.7778 & 0.7917 \\
  (6, 2)   & 0.7912 & 0.7976 & 0.8052 & 0.8143 \\
  (7, 2)   & 0.8167 & 0.8214 & 0.8269 & 0.8333 \\
  (4, 3)   & 0.6564 & 0.6981 & 0.7516 & 0.8231 \\
  (5, 3)   & 0.7102 & 0.7375 & 0.7708 & 0.8125 \\
  (6, 3)   & 0.7496 & 0.7688 & 0.7915 & 0.8188 \\
  (7, 3)   & 0.7795 & 0.7938 & 0.8103 & 0.8295 \\
  \hline  
\end{tabular}
\end{table}

\begin{table}
\caption{Comparison of upper and lower bounds on the rate of $(r,t,x)$-LRC codes}
\label{t_2}
\centering
\begin{tabular}{| l || c | c || c | c | c |}
  \hline                       
  $(r, t)$ &  $x=0$ (WZL) & $R^*(r,t, x=0)$ & $x=1$ & $R^*(r,t, x=1)$  \\
  \hline
  (3, 2)   & 0.6000 & 0.6429 & 0.6667 & 0.6667 \\
  (5, 2)   & 0.7143 & 0.7576 & 0.7500 & 0.7667 \\
  (7, 2)   & 0.7778 & 0.8167 & 0.8000 & 0.8214 \\
  (3, 3)   & 0.5000 & 0.5786 & 0.6250 & 0.6500 \\
  (5, 3)   &  0.6250 & 0.7102 & 0.7000 & 0.7375 \\
  (7, 3)   &  0.7000 &  0.7795 & 0.7500 & 0.7938 \\
  \hline  
\end{tabular}
\end{table}

\section{Conclusion}
We investigated one possible generalization of locally recoverable codes (LRC) with all-symbol locality and availability when recovering sets can intersect in a small number of coordinates. This feature allows us to increase the achievable code rate and still meet load balancing requirements. In this paper we derived an upper bound for the rate of such codes and gave explicit constructions of codes with such a property.

\section*{Acknowledgment}
A.~Frolov thanks A.~Barg for introducing this problem to him and for numerous fruitful discussions during his stay in University of Maryland.

\end{document}